\documentstyle[epsfig,12pt]{article}

\newlength{\dinwidth}
\newlength{\dinmargin}
\def\be{\begin{eqnarray}}
\def\ee{\end{eqnarray}}

\def\d{{\rm d}}

\newcommand{\gsim}{\mathrel{\mathop{\kern 0pt \rlap
  {\raise.2ex\hbox{$>$}}}
  \lower.9ex\hbox{\kern-.190em $\sim$}}}
\newcommand{\lsim}{\mathrel{\mathop{\kern 0pt \rlap
  {\raise.2ex\hbox{$<$}}}
  \lower.9ex\hbox{\kern-.190em $\sim$}}}

\begin{document}
\rightline{SNUST030302} \rightline{ITFA-2003-14}
\rightline{KIAS-P03022} \rightline{\tt hep-th/0303235}
\vspace*{0.5cm} \centerline{\Large \bf Softer Hard Scattering}
\vspace*{0.3cm} \centerline{\Large \bf and} \vspace*{0.3cm}
\centerline{\Large \bf Noncommutative Gauge--String Duality
\footnote{Work supported in part by the KRF Overseas Research
Grant, the KOSEF Interdisciplinary Research Grants, the KOSEF
Leading Scientist Grant, and the Stichting FOM.}}\vspace*{1.0cm}
\centerline{\bf Soo-Jong Rey ${}^{a,b}$ {\rm and} \bf Jung-Tay Yee
${}^{c,d}$} \vspace*{1.0cm}
\centerline{\sl School of Natural Sciences, Institute for Advanced
Study} \vspace*{0.2cm} \centerline{\sl Einstein Drive, Princeton
NJ 08540 \rm USA ${}^a$}
\vspace*{0.4cm}
\centerline{\sl School of Physics \& BK-21 Physics Division}
\vspace*{0.2cm} \centerline{\sl Seoul National University, Seoul
151-747 \rm KOREA ${}^b$}
\vspace*{0.4cm}
\centerline{\sl Institute for Theoretical Physics, University of Amsterdam} %
\vspace*{0.2cm}
\centerline{\sl Valckenierstraat 65, 1018 XE Amsterdam,  \rm The
NETHERLANDS ${}^c$}
\vspace*{0.4cm}
\centerline{\sl School of Physics, Korea Institute for Advanced
Study} \vspace*{0.2cm} \centerline{\sl Seoul 130-012 \rm KOREA
${}^d$}
\vspace*{0.8cm}
\centerline{\tt sjrey@gravity.snu.ac.kr \hskip1cm
jungtay@science.uva.nl}
\vspace*{0.8cm}
\begin{abstract}
We study exclusive scattering of `hadrons' at high energy and
fixed angle in (nonconformal) noncommutative gauge theories. Via
gauge-string duality, we show that the noncommutativity renders
the scattering soft, leading to exponential suppression. The
result fits with the picture that, in noncommutative gauge theory,
`fundamental partons' consist of extremely soft constituents and
hadrons are made out of open Wilson lines.

\end{abstract}

\baselineskip=18pt
\newpage

In this paper, we pose the question: ``{\sl what are the
fundamental degrees of freedom in noncommutative gauge theory?}''
and find an answer to it.  Study of noncommutative gauge theory
\cite{connesdouglasschwartz, seibergwitten}, motivated prominently
because the theory describes D-brane worldvolume dynamics under
B-field background in string theory, has brought us many
surprises. First, what one might counted as an ultraviolet (UV)
effect is sometimes transmuted to an infrared (IR) effect.
Diagrammatically, origin of this so-called UV-IR mixing
\cite{uvir} is quite reminiscent of the channel duality between
open and closed strings. Second, the quantum dynamics is
describable entirely in terms of open Wilson lines \cite{loop}
(see also \cite{others}). The open Wilson lines \cite{reyvonunge}
are nonlocal, gauge-invariant operators, and behave as a dipole
under a strong magnetic field, analogous to fundamental strings.
Both features point to the possibility that partons in the
noncommutative gauge theory are not pointlike gluons but some
extended objects. We will attempt to answer the question posed by
utilizing the gauge-string duality, and by examining exclusive
processes such as high-energy, fixed-angle scattering
\cite{brodskyfarrar}.

For commutative gauge theory, Polchinski and Strassler
\cite{polstr} recasted the fixed-angle scattering of glueballs
entirely in terms of gauge-string duality \cite{maldacena}, and
have shown that the known power-like behavior \cite{brodskyfarrar}
at high-energy regime is retractible. In terms of the gauge-string
duality, the glueball scattering in the gauge theory side is
described by the dilaton scattering in the bulk string theory
side. At high-energy, as the latter process is described by the
Virasoro-Shapiro amplitude, one might naively expect that the
scattering process is exponentially suppressed.

The above expectation turns out incorrect \cite{polstr}. By the
holography principle in anti-de Sitter space and the superposition
principle in quantum mechanics, the scattering amplitude in gauge
theory is given by a coherent sum of the string scattering
amplitude over the bulk location where the scattering takes place.
For this, because of the holographic UV-IR relation
\cite{quarkpot} in the gauge-string duality, one finds that the
sum is dominated by the process taking place near the boundary.
Key feature here is that, because of the warp factor, the local
inertial momentum $\widetilde{\bf p}$ is mapped to the conserved,
gauge theory momentum ${\bf p}$ via $\widetilde{\bf p}_m = \left(R
\slash \ell_{\rm st} u \right) {\bf p}_m$, so, for any value of
${\bf p}$, one can locate a bulk position $u$, where
$\widetilde{\bf p}$ becomes ${\cal O}(1/\ell_{\rm st})$, leading
to ${\cal O}(1)$ contribution of the string amplitude. This
explains why the glueball scattering as deduced via the
gauge-string duality actually behaves power-like at high-energy.
It provokes one to inquire if the elastic scattering processes are
revealing point-like, hard parton nature of the fundamental
degrees of freedom in commutative gauge theory.

From the gauge theory viewpoint, one can portray the question
posed from a different angle: "is it possible to find a
\underline{field-theoretic} deformation of commutative gauge
theory so that low-energy dynamics remains unmodified but
high-energy parton content is dramatically modified?" The answer
is affirmatively yes, and is achieved by defining the theory on a
noncommutative space-time. Our result would be considered as a
{\sl direct} evidence for the assertion.

For ${\cal N}=4$ supersymmetric gauge theory with noncommutativity
on $(23)$-plane turned on, the background fields in the dual
string theory is given by \cite{malrus}
\be \label{metric}
 ds^2 &=& \ell^2_{\rm st} \left[ {u^2 \over R^2} \left\{ - dx_0^2 + dx_1^2 +
  {1 \over 1+  a^4 u^4 } \left(dx_2^2 + dx_3^2 \right) \right\}
  + {R^2 \over u^2 } du^2 + R^2 d\Omega_5^2 \right] \nonumber \\
  e^{2 \phi} &=& g_{\rm st}^2 \left( 1 + a^4 u^4
  \right)^{-1}
  \qquad \qquad
  G_{0123u} = {1 \over g_{\rm st}}
  \partial_u \left( {\ell^4_{\rm st} \over R^4} u^4 \right)
  \left(1 + a^4 u^4 \right)^{-1}
  \nonumber \\
  B_{23}^{\rm NS} &=& {\ell^2_{\rm st} \over \Theta}
  {\Theta^2 \over R^4} u^4 \left( 1+ a^4 u^4
  \right)^{-1}
  \qquad \qquad
  C_{01}^{\rm RR} = {1 \over g_{st}} {\ell^2_{\rm st} \over \Theta}
  a^4 u^4.
  \ee
Here, the curvature radius is given by $R \ell_{\rm st}$, where
$R$ is related to the string coupling parameter and number of
colors by $R= \left(4\pi g_{\rm st} N \right)^{1/4}$. The
parameter $a^4 = \Theta^2 / R^4$ sets the noncommutative scale. In
the coordinates adopted, the noncommutativity is denoted by
$\left[ \widehat{x_2}, \widehat{x_3} \right]= i \Theta$. As we are
interested primarily in studying how the noncommutativity affects
high-energy fixed-angle scattering processes, we will restrict the
kinematics so that the scattering takes place in the
noncommutative subspace. The momentum $\widetilde{\bf p}$ in local
inertial coordinates and the conserved momentum $\bf p$ at the
holographic boundary are then related by the geometry
Eq.(\ref{metric}) as
\be \label{uvir}
 \widetilde{\bf p}_m =
{e_m}^{\underline{m}} {\bf p}_{\underline{m}} =
 {\sqrt{\Theta} \over \ell_{\rm st}} \left( {R^2 \slash \Theta \over u^2} +
 {u^2 \over R^2 \slash \Theta} \right)^{1/2} \!\!{\bf p}_m . \ee
In the commutative regime, $u \ll R \slash \sqrt{\Theta}$, and the
relation reduces to the same as in the conformal gauge theory
\cite{quarkpot} $\widetilde{\bf p}_m \sim \left( R \slash
\ell_{\rm st} u \right) {\bf p}_m$.
As such, for a high gauge-momentum ${\bf p}$, it was always
possible to find a region in the bulk, where the string-momentum
$\widetilde{\bf p}$ is of order unity. On the other hand, in the
noncommutative regime, $u \gg R \slash \sqrt{\Theta}$, and the
relation is given by $
 \widetilde{\bf p}_m \sim \left( \Theta u \slash \ell_{\rm st} R
 \right) {\bf p}$.
 Thus, for a fixed string-momentum $\widetilde{\bf p}$,
 the gauge-momentum$\bf p$ is \underline{inversely}
 proportional to $u$, precisely the opposite to the commutative
 regime.

So, what makes the story so interesting for noncommutative gauge
theory is that, for the very high gauge-momentum $\bf p$, there is
no $u$-region in the bulk where the inertial string-momentum
$\tilde{\bf p}$ can be made small enough
--- the conversion factor in Eq.(\ref{uvir}) is bounded below.
This will become the key reason why noncommutative gauge theory
behaves string-like at high energy. For the scattering at high
enough energy, one cannot adjust $u$-location of the scattering
center so that $\tilde{\bf p}$ remains small. The local inertial
string-momentum $\widetilde{\bf p}$ has to grow in proportion to
the gauge-momentum ${\bf p}$. Thus, up to a finite rescaling,
high-energy processes in gauge theory corresponds to high-energy
processes in string theory as well. We will see momentarily that
the string theory processes take place at the location $u \sim R
\slash \sqrt{\Theta}$. Notice that, at this location, the
proportionality factor in Eq.(\ref{uvir}) beomces
\underline{independent} of the `t Hooft coupling parameter.

We now investigate the high-energy, fixed-angle limit of exclusive
 scalar glueball scattering, and compare the result with the one
for commutative gauge theory. In exclusive processes, in- and
out-states are created by gauge-invariant operators. In
noncommutative gauge theory, this requires adjoining the
leading-twist glueball operator $F_{mn}^2$ with open Wilson lines
\cite{reyvonunge}. The simplest exclusive process would be the
elastic two-body scattering, and the scattering amplitude in gauge
theory is obtainable by convoluting the corresponding string
amplitude over the wave function of the bulk state, e.g. the
dilaton for scalar glueballs. Thus,
\be \label{amp}
 {\cal A}_{\rm NCYM}({\bf p}) = \int_0^\infty \d u \int_{S_5} \d \Omega_5
 \sqrt{g_{(10)}} \, {\cal A}_{\rm string}(\widetilde{\bf p}, \Omega_5)
\prod_{i=1}^4 \Psi_i(\widetilde{\bf p}_i, u),
 \ee
where $\Psi(\widetilde{\bf p}, u)$ is the normalized wave-function
of the dilaton field in the bulk, and $g_{(10)}$ is the volume
factor of the ten-dimensional space-time (in type IIB string
frame). From Eq.(\ref{metric}), $
 \sqrt{g_{(10)}}= \ell_{\rm st}^{10}
  R^2 u^3 \left( 1+ {\Theta^2 u^4 \slash R^4}
 \right)^{-1}$.
The string scattering process taking place in the bulk is
complicated as background fields -- metric, dilaton, NS-NS 2-form,
and R-R 2- and 5-forms -- are turned on. In particular,
contribution of the R-R field background is of the same order as
that of the NS-NS fields, since the former has strength of order
$1/g_{\rm st}$. With currently available methods, analytic
computation of string scattering amplitude in such situation is
impossible. Nevertheless, in the limit $R \rightarrow \infty$, the
background is reduced to the one with flat space-time, constant
dilaton, and vanishing NS-NS and R-R field strengths. Thus, at
leading-order in $1/R$, the string scattering amplitude can be
approximated by the Virasoro-Shapiro amplitude.

Consider a {\sl large} but {\sl fixed} gauge-momentum ${\bf p}_m$.
When converted into the inertial string-momentum $\widetilde{\bf
p}_m$, the conversion factor depends on the $u$-coordinate. At the
minimum of the conversion factor, {\sl both} the mean-value and
the variance of the $u$-coordinate is set by $R \slash
\sqrt{\Theta}$. Thus, by letting the noncommutativity $\Theta
\slash \ell_{\rm st}^2 \equiv \mu^2 \gg 1$ but $R \slash \mu \sim
{\cal O}(1)$, one can always find a situation that the scattering
process is pinned at a fixed $u$-position and the wave function is
spread out typically only over the string scale $\ell_{\rm st}$.
In such a case, combined with the argument given in the previous
paragraph, we can adopt the Virasoro-Shapiro amplitude, which
refers to the scattering process in a flat ten-dimensional
background, as the leading-order string scattering amplitude. We
recall that the Virasoro-Shapiro amplitude is given by
\be
 {\cal A}_{\rm string}(\tilde{{\bf p}})= g_s^2 \ell_{\rm st}^6  {\Gamma
 \left(-{1 \over 4} \ell_{\rm st}^2 \tilde{s} \right)
    \Gamma \left(-{1 \over 4} \ell_{\rm st}^2 \tilde{t} \right)
    \Gamma \left(-{1 \over 4} \ell_{\rm st}^2 \tilde{u} \right)
     \over \Gamma \left(1+{1 \over 4} \ell_{\rm st}^2 \tilde{s} \right)
    \Gamma \left(1+{1 \over 4} \ell_{\rm st}^2 \tilde{t} \right)
    \Gamma \left(1+{1 \over 4} \ell_{\rm st}^2 \tilde{u} \right) }
   K_c,
\nonumber \ee
where $K_c$ is a kinematical factor. For high-energy and
fixed-angle scattering, $\widetilde{s} \rightarrow \infty$ and
$\widetilde{t} \rightarrow \infty$ with
$\widetilde{s}/\widetilde{t}$ held fixed, and the string
scattering amplitude exhibits a soft behavior
\be \label{bulkamp} {\cal A}_{\rm string}(\widetilde{\bf p}) \sim
g_s^2 \ell_{\rm st}^6 \left(\ell_{\rm st}^2 \tilde{s}\right)^3
\sin^2 \theta \left[ \exp\left({f(\theta) \over
2}\right)\right]^{-\ell_{\rm st}^2 \tilde{s}}. \ee
Here, we have used the mass-shell condition
$\tilde{s}+\tilde{t}+\tilde{u}=0$, and defined the scattering
angle $\theta$ as $\tilde{t}=-\tilde{s} \sin^2{\theta \over 2}$
and $ \tilde{u}=-\tilde{s}\cos^2{\theta \over 2}$. The parametric
function $f(\theta)= -\sin^2 {\theta \over 2} \ln \left(\sin^2
{\theta \over 2} \right)
 - \cos^2 {\theta \over 2} \ln \left( \cos^2{\theta \over 2} \right) $
is a positive-definite function of $\theta$ in the entire range
$[0, \pi]$.

What becomes modified dominantly by turning on the
noncommutativity is the normalized wave-function of the bulk
scalar field, $\Psi(u, \Omega_5)$. The wave-function was computed
already in \cite{reyvonunge, dasgho}, and is given by $
 \Psi(u, {\bf p}) \sim u^{-2} e^{-i {\pi \over 2}(\nu+1)}
          H^{(2)} \left(\nu, -\log( \Theta u)
         \right),$
where $H^{(2)}$ denotes the second kind of the Mathieu function,
 and $\nu$ refers to so-called Floquet coefficient (which depends
 implicitly on the momentum ${\bf p}$). For foregoing
 analysis, we have found it sufficient and more illuminating to
 understand asymptotic behavior of the normalized wave-function,
 $\psi(u, \Omega_5)$. The wave function obeys the dilaton
 equation of motion: $\partial_M \left( \sqrt{g_{(10)}} g^{MN}_{(10)}
 e^{-2 \phi(u)} \partial_N \Psi \right)(x, u, \Omega_5) =0$.
Fourier transforming with respect to $x^m$, taking S-wave mode on
$S_5$, and rescaling as $\Psi(u, {\bf p}) = u^{-{5 \over 2}}
\psi(u, {\bf p})$, we recast the equation of motion into a form of
the Schr\"{o}dinger equation:
\be -\partial_u^2 \psi (u, {\bf p}) + {\cal K}^2(u, {\bf p}) \psi
(u, {\bf p}) = 0 \qquad {\rm where} \qquad {\cal K}^2(u, {\bf p})
= \left( {R^4 {\bf p}^2 \over u^4} + {15\over 4} {1 \over u^2} +
|{\bf p} \Theta|^2 \right). \label{eom} \ee
Classical turning point $u = u_{\rm c}$ of Eq.(\ref{eom}) is
determined by ${\cal K}(u_{\rm c}, {\bf p}) = 0$. At high-energy
limit $\vert {\bf p} \vert \rightarrow \infty$ we are interested
in, it turns out the turning point is located at $ u_c \sim  u_* =
{R / \sqrt{\Theta}}$, and this is precisely the scale set by the
noncommutativity, as seen in Eq.(\ref{uvir}). Asymptotic form of
the wave function is obtainable via the WKB method. The result is
\be \label{wkb}
 \Psi(u, {\bf p}) \,\, \simeq \,\,
 \left\{ \begin{array}{lll}
 C({\bf p}) \, |{\bf p} \Theta|^{-1/2} u^{-5/2}
 e^{-|{\bf p}\Theta| (u-u_*) }
 & \qquad\quad  & u \gg u_* \\
 && \\
 2 C({\bf p}) \,\vert {\bf p} R^2\vert^{-1/2} u^{-3/2}  \cos\left( \vert{\bf p} R^2 \vert
 \left({1 \over u} - {1 \over u_*}\right) -{\pi \over 4} \right)
 & \qquad\quad  & u \ll u_*\\
 &&\\
 2 \sqrt{\pi} C({\bf p}) \, u^{-5/2} h^{-{1 \over 6}}({\bf p}) \,Ai \left(h^{{1\over
 3}}({\bf p})   (u-u_*) \right)
 & \qquad\quad  & u \approx u_* .
 \end{array} \right.
\ee
Here, $C({\bf p})$ is a momentum-dependent normalization constant,
$h({\bf p})$ is defined as $h({\bf p}) \equiv [\partial_{u} {\cal
K}^2 (u, {\bf p}) ]_{u=u_*}$, and $Ai(x)$ refers to the Airy
function. In the high-energy limit, the coefficient function
$h({\bf p})$ is approximated as $h({\bf p}) \approx 4 {\bf p}^2
\Theta^{5/2}/R $. The normalization constant $C({\bf p})$ is
determined by requiring that the wave-function obeys correctly
normalized commutation relations, and this results in the
condition: $\int_0^\infty \d u \, g^{00}_{(10)} \sqrt{g_{(10)}} \,
|\Psi|^2 = 1$. As $|\Psi|(u, {\bf p})$ in Eq.(\ref{wkb}) is
exponentially damped for large $u$ and rapidly oscillatory for
small $u$, the normalization integral is dominated by the region
near $u \approx u_*$. Evaluating the integral by the saddle-point
method, we determined $C({\bf p})$ in terms of the coefficient
function $h({\bf p})$ as $|C({\bf p}) h^{-{1 \over 6}}({\bf p})|^2
\sim \left( u_*^3 \slash \ell_{\rm st}^8
 R^4 \right)$.

It now remains to collect all parts and put them together.
Re-expressing all momentum dependence in terms of gauge-momentum
Eq.(\ref{uvir}), we obtain the string scattering amplitude
Eq.(\ref{bulkamp}) in the high-energy limit as
\be \label{bulkscatt}
 {\cal A}_{\rm string} ({\bf p}) \sim g_{\rm st}^2
 \ell_{\rm st}^6 (\Theta s)^3 \sin^2 \theta
 \left( {u_*^2 \over u^2}
  + {u^2 \over u_*^2} \right)^3  \left[ \exp\left( {f(\theta) \over 2}
   \right) \right]^{- \left( {u_*^2 \over u^2}
  + {u^2 \over u_*^2} \right) \Theta s }.
\ee
With the above WKB results, the convolution integral
Eq.(\ref{amp}) is not doable analytically, so we shall evaluate it
via the saddle-point method. For large and small values of $u$,
the integrand in Eq.(\ref{amp}) is exponentially suppressed, as we
see from Eq.(\ref{bulkscatt}). The suppression is dominantly due
to the string scattering amplitude Eq.(\ref{bulkscatt}), and the
normalized dilaton wave function renders further, sub-leading
suppression at large $u$. Thus, leading contribution to the
convolution integral Eq.(\ref{amp}) comes from the region around
the saddle-point $u=u_*$. Near the saddle-point, the Airy function
yields ${\cal O}(1)$ contribution. Consequently, we find that
exponentially soft behavior dominates the gauge theory amplitude
at high-energy:
\be {\cal A}_{\rm NCYM}({\bf p}) \sim { (g_{\rm YM}^2 N)^{1 \over
2} \over N^2}
  (\Theta s)^3 \sin^2 \theta \left[
e^{f(\theta)} \right]^{-\Theta s} \qquad {\rm for} \quad \vert
{\bf p} \vert \rightarrow \infty, \label{final} \ee
where subdominant terms are omitted. Notice that, in
Eq.(\ref{final}), dependence on the string scale $\ell_{\rm st}$
is cancelled out while dependence on the gauge theory coupling
parameters is identical to the commutative counterpart
\cite{polstr}, as anticipated. As Eq.(\ref{final}) refers to the
leading-order result at large noncommutativity, $\Theta \gg
\ell^2_{\rm st}$, naive commutative limit of Eq.(\ref{final}) is
not permitted. Redoing the saddle-point analysis at finite
$\Theta$, however, one readily finds that Eq.(\ref{final})
interpolates smoothly to the commutative result of \cite{polstr}:
from Eq.(\ref{final}), the exponential suppression disappears, and
the pre-exponential part is replaced by a constant (up to
sub-leading corrections).

Eq.(\ref{final}) is the main result of this paper. What is most
remarkable of the result Eq.(\ref{final}) is that, in
noncommutative gauge theory, the exclusive scattering amplitude is
extremely soft at high-energy, damped exponentially. In fact, the
scattering amplitude is the same as that of the string theory
except the role of the string scale $\ell_{\rm st}$ is now played
by the noncommutative scale $\sqrt{\Theta}$. The stringy behavior
is pronounced most for the energy-momentum exceeding the value set
by the noncommutativity scale: $\vert {\bf p} \vert \gsim {1
\slash \sqrt{\Theta}}$.
According to the gauge-string duality, the dilaton scattering in
the string theory is described by the glueball scattering in the
gauge theory. In noncommutative gauge theory, it is known that the
interpolating operator of the glueball is not just $F^2_{mn}(x)$
but the one adjoined by the open Wilson line \cite{reyvonunge}.
The size ${\bf d}$ of the open Wilson line grows linearly with the
center-of-mass momentum ${\bf p}$: ${\bf d} = \Theta \wedge {\bf
p}$. It then implies that, in noncommutative gauge theory, stringy
behavior of the partons are manifested by the open Wilson lines,
whose size grows larger at stronger noncommutativity: $\vert {\bf
d} \vert \gsim \sqrt{\Theta} $.
As such, `hadrons' adjoined with such large open Wilson lines are
interpretable as a sort of `closed string' modes emergent in
noncommutative gauge theory.

In \cite{reyvonunge}, Gross {et.al.} also reported soft behavior
of high-energy scattering in noncommutative gauge theory. There,
however, despite apparent similarity of the result with ours, the
exponential suppression bears no connection to the exponential
suppression in the dual string scattering amplitude. This is
because their exponential suppression originates from the
bulk-to-boundary propagator ($u \gg u_*$ region of Eq.(\ref{wkb}))
and was entirely within the supergravity approximation, under
which no stringy features would remain. In contrast, in our result
Eq.(\ref{final}), exponential suppression originates dominantly
from the full-fledged string scattering amplitude ${\cal A}_{\rm
string}$ (dilaton wave function served mainly to provide the
requisite kinematical parameters and coupling constant of the dual
gauge theory). We emphasize importance of string theoretic
computation in the bulk. Had we taken a supergravity approximation
and replaced Eq.(\ref{bulkamp}) by the four-point dilaton
scattering amplitude, the result would come out glossly different
and incorrect at high-energy. For that reason, in our result,
correspondence of exponential suppression between the string
theory and the noncommutative gauge theory is exact.

Obviously, next task is to compute high-energy scattering
amplitude directly from the noncommutative gauge theory and to
confirm the stringy behavior for the scattering among four
`glueball' operators involving open Wilson lines. In
\cite{dharkitazawa}, such processes were studied, but exclusively
at leading order at weak coupling regime. An interesting issue for
future study is to establish a suitable interpolation from weak to
strong coupling regime, and relate to our result.

We showed explicitly, in terms of gauge-string duality, how
`closed string'-like behavior appears in noncommutative gauge
theory. Though, from purely field theoretical consideration, the
appearance of `closed string' behavior in noncommutative field
theories is rather surprising, we have learned that exclusive
high-energy processes offers an ideal set-up for answering the
reason why via the gauge-string duality. A further hint may be
learned from the full-figured ten-dimensional geometry of D3-brane
before taking the near-horizon limit. The geometry is claimed
\cite{uvflow} to be dual to a UV-flow of the four-dimensional
${\cal N}=4$ supersymmetric gauge theory once special irrelevant
operators are turned on. If, tentatively, the holographic boundary
is placed at asymptotic infinity (which is the ten-dimensional
flat space-time), one obtains the local inertial momentum ${\bf
p}$ is related to the gauge theory momentum ${\bf p}$ as
$\widetilde{\bf p} = \sqrt{1+ {\ell^4_{\rm st} R^4 \slash u^4}}
{\bf p}$, we see that the resulting relation is qualitatively
similar to Eq.(\ref{uvir}): if we take large $u \gg 1/(R \ell_{\rm
st})$, $\widetilde{\bf p}$ is proportional to ${\bf p}$. This
means that closed string modes do survive, and the holographic
theory would behave precisely like IIB string theory in ten
dimensions.

Summarizing our results, we have demonstrated that noncommutative
gauge theory exhibits dramatically modified elastic scattering
behavior from commutative counterparts. At energy-momentum
transfer well below the noncommutativity scale, both theories are
indistinguishable. As energy-momentum transfer is increased toward
the noncommutativity scale, both theories exhibit parton-like
elastic scattering behavior. Underlying to the behavior is
attributable to the kinematics of conformal invariance at energy
scale far above any nonperturbative scales and not to underlying
dynamics of partons. At energy-momentum transfer well above the
noncommutativity scale, the two theories display dramatically
different elastic scattering behavior --- while commutative gauge
theory continue exhibiting parton-like power-law behavior,
noncommutative gauge theory begins to expose nonlocal and
string-like constituents. As the behavior originates from
departure of the bulk asymptotic geometry from anti-de Sitter and
as the departure is precisely the feature underlying emergence of
dipole-like excitations at very high-energy scale, one may
interpret quanta created by the open Wilson lines as the
fundamental constituents of the noncommutative gauge theory.

More recently, Polchinski and Strassler \cite{Polchinski2}
extended the gauge-string duality for commutative gauge theory to
deep inelastic scattering processes at large `t Hooft coupling,
and found that evolution of the structure functions is more rapid
than those at small `t Hooft coupling. This implies that there are
no hard-momentum carrying partons inside hadrons but only wee
partons. In light of the result of this work, it would be quite
interesting to extend the study of \cite{Polchinski2} to
noncommutative gauge theory and examine if high-energy evolution
of the structure functions is modified further than the wee-parton
behavior beyond the noncommutativity
scale.
\section*{Acknowledgement}
 We thank D.J.
Gross, J. Polchinski, L. Susskind, and C. Thorn for interesting
conversations during the "QCD and String Theory" Workshop at the
Institute for Nuclear Theory at Seattle. We also thank the
anonymous referee for important remark concerning interpretation
of the results of \cite{polstr, Polchinski2}. This work was
carried out while SJR was a Member at the Institute for Advanced
Study. He thanks the School of Natural Sciences for hospitality
and the Funds of Natural Sciences.

\end{document}